\documentclass[12pt]{article}

\usepackage{amsfonts,amsmath,amsxtra}

\newcommand{\beq}[1]{\begin{equation}\label{#1}}
\newcommand\eeq {\end{equation}}
\newcommand\bqa {\begin{eqnarray}}
\newcommand\eqa {\end{eqnarray}}

\newcommand{\eq}[1]{eq.(\ref{#1})}
\newcommand{\bear}{\begin{array}}
\newcommand{\enar}{\end{array}}

\begin{document}

\def\t{\theta}
\def\T{\Theta}
\def\w{\omega}
\def\ov{\overline}
\def\a{\alpha}
\def\b{\beta}
\def\g{\gamma}
\def\s{\sigma}
\def\l{\lambda}
\def\wt{\widetilde}


\vspace{10mm}

\centerline{\bf \Large On the physical meaning of the Unruh
effect}

\vspace{7mm}

\centerline{{\bf Emil T.Akhmedov}\footnote{akhmedov@itep.ru}}
\centerline{Moscow, B.Cheremushkinskaya, 25, ITEP, Russia 117218}
\vspace{3mm} \centerline{and} \vspace{3mm} \centerline{{\bf
Douglas Singleton}\footnote{dougs@csufresno.edu}}
\centerline{Physics Department, CSU Fresno, Fresno, CA 93740-8031}
\vspace{10mm}

\begin{abstract}
We present simple arguments that detectors moving with constant
acceleration (even acceleration for a finite time) should detect
particles. The effect is seen to be universal. Moreover, detectors
undergoing linear acceleration and uniform, circular motion both
detect particles for the same physical reason. We show that if one
uses a circularly orbiting electron in a constant external
magnetic field as the Unruh--DeWitt detector, then the Unruh
effect physically coincides with the experimentally verified
Sokolov--Ternov effect.
\end{abstract}

\vspace{10mm}

Hawking radiation \cite{Hawking} and the closely related Unruh
\cite{Unruh} radiation are often seen as first steps toward
combining general relativity and quantum mechanics. Under
achievable conditions for gravitational system these effects are
too small to be experimentally testable. In this letter we examine
the physical meaning of the Unruh effect and argue that for
uniform, circular acceleration the Unruh effect has already been
observed. Given the close connection between the Hawking and Unruh
effects this experimental evidence for the latter gives strong
support for the former.

It has been shown \cite{Unruh} that a detector moving eternally
with constant, linear acceleration $a$ should detect particles
with Planckian distribution of temperature $T=a/2\pi$. The
non--inertial reference frame which is co--moving with the
detector has an event horizon. Even massless particles radiated a
distance $1/a$ behind the detector would never catch up with an
eternally accelerating detector. It is the reference frame
co--moving with the {\it eternally} accelerating detector which
``sees'' the Rindler metric. Thus it seems that the Unruh effect
is strongly related to the existence of the horizon. However, if
the effect only exists for an eternally accelerating
observer/detector then it can be discarded as unphysical since one
can never have a detector that undergoes constant acceleration
from infinite past time to infinite future time. Due to the
Hawking radiation \cite{Hawking} black-holes do not exist
eternally. As well a positive cosmological constant (giving a
de-Sitter space--time) should eventually be radiated away to zero.

The real question is whether or not a detector which moves with
linear, constant acceleration for a finite time will see particles
(e.g. a detector which is initially stationary, accelerates for a
finite time and then continues with constant velocity). We are
interested whether the detector gets excited or not during the
period when it moves homogeneously. We are not interested in the
detector's reaction during the periods when the acceleration is
turned on or off. The reaction of the detector which we are
interested in does not come from internal forces where one part of
the detector can move with respect to another (like the arrow of
an ammeter which moves with respect to its box if it is shaken),
but is due to the existence of a universal radiation in the
detector's non--inertial reference frame. We consider two kinds of
homogeneous accelerations: (i) from a force that is constant in
magnitude and direction resulting in linear accelerated motion;
(ii) from a force that is constant only  in magnitude resulting in
circular motion. We take as our definition of a particle that
thing which causes a detector to click, i.e. jump from one of its
internal energy levels to a higher one. We do not know any other
{\it invariant} definition of a particle.

If detectors do click during homogeneous, accelerated motion
occurring for a finite time, then the Unruh effect does not depend
on the existence of a horizon\footnote{Here we understand notion
of the horizon as the {\it eternally} existing surface from inside
of which classically nothing can {\it ever} escape.}, since for
finite time acceleration the co--moving frame ``sees'' a metric
different from Rindler and does not have a horizon: a massless
particle with light speed velocity following the detector will
eventually catch up with it if the detector accelerates for finite
time.

Once this idea is accepted, we can go further and state that there
is no significant physical difference between detectors in
homogeneous, linear acceleration versus uniform circular motion.
Note, the reference frame co--moving with the detector performing
eternal homogeneous, circular motion does not have a horizon (only
a light--surface). A particle can eventually catch up with a
circularly moving detector. Previous investigations on whether or not
a moving detector clicks or not under various assumptions about
its motion can be found in references \cite{padman} \cite{louko}

In this letter we show (following other authors) that detectors
performing homogenous linear and circular accelerations (or any
other homogeneous non--inertial motion in the empty Minkowski
space) do detect particles, and they do this for the same physical
reason. Moreover, we show that the circular Unruh effect has been
well known for a long time under a different name and has even
been experimentally observed.

In all cases we consider Minkowski space--time, and take $\hbar
=1$ and $c=1$. For simplicity we consider a linear interaction of
the detector with a {\it free} scalar field. We consider the
following two processes: (i) the detector is originally in its
ground state and then gets excited because of its non--inertial
motion; (ii) the detector is originally in its excited state and
then relaxes to its ground state. In both cases the background QFT
is originally in its ground state. We want to find the probability
rates for these two processes. As a result of these processes the
background QFT will become excited, i.e. the detector will radiate
quanta of the background QFT when performing the above two
processes.

To leading order in perturbation theory the probability rate per
unit time is \cite{BirelDavis}: \bqa w_{\mp} \propto
\int_{-\infty}^{+\infty} d\tau \, e^{\mp {\rm i}\, \Delta
\mathcal{E}\, \tau}\, G\left[x(t -
\tau/2)^{\phantom{\frac12}},\,\,x(t + \tau/2)\right],
\label{grule2} \eqa where $t$ is the detector's proper time;
$\Delta \mathcal{E} = \mathcal{E}_{up} - \mathcal{E}_{down} > 0$
is the discrete change of the detector's internal energy level;
the ``$-$'' sign, both in the LHS and in the exponent, corresponds
to the first process, while the ``$+$'' sign corresponds to the
second process mentioned above; $G\left[x(t - \tau/2),\,x(t +
\tau/2)\right] = \left\langle 0\left|\phi[x(t
 - \tau/2)]^{\phantom{\frac12}} \phi[x(t + \tau/2)]\right| 0\right\rangle$ is
the Wightman function of the scalar field $\phi$. This function
measures the correlation between fluctuations of the scalar field
at two points in the space--time in the vacuum of the scalar QFT.
In our case these two points are on the same trajectory $x(t)$.
Because of this these points are causally connected to each other
even for the eternally, linearly accelerating detector. However,
as we will see below the important contribution to $w_{\mp}$ in
all cases comes from the imaginary $\tau$.

The reason why we consider the detector approach to the Unruh
effect is that then all our considerations can be made completely
generally covariant \cite{AkhSing3}. This allows us to address the
question as to whether or not a detector making a particular
motion in Minkowski space--time sees/detects particles.

Eq. (\ref{grule2}) shows that the probability rates $w_{\mp}$ are
Fourier images of the Wightman function. The Wightman function is
a universal characteristic of the field, and its features {\it
universally} characterize the reaction of a detector moving along
the trajectory $x(t)$. Of course the spectrum of the detected
particles depends on the detector's trajectory.

Note that \eq{grule2} is written for the simplest linear type of
interaction of the detector with $\phi$ \cite{BirelDavis}
\cite{AkhSing3}. In cases with a more complicated interaction, say
non--linear or via derivatives of the field, one would get
probability rates that are Fourier images of powers or derivatives
of the Wightman function. It will be clear from the discussion
below that this would not change the spectrum of the detected
particles, but would only alter the time necessary to reach the
equilibrium distribution over the detector's energy levels under
the homogeneous background radiation.

Thus, the question is reduced to the study of the characteristic
features of the Wightman function of free massless particles: \bqa
G(x,y) = \frac{1}{\left|x_0 - y_0 - {\rm i}\, \epsilon\right|^2 -
\left|\vec{x} - \vec{y}\right|^2}, \label{explic} \eqa with
various homogeneous trajectories -- $x(t_1) = x$ and $x(t_2) = y$
 -- plugged into it. Below we are going to consider three different
trajectories. All poles of the two--point correlation functions
(both in coordinate and momentum spaces) have physical meanings
based on intuition from condensed matter physics.

In the case of motion with constant velocity one can show that
(see e.g. \cite{AkhSing3}): $w_- = 0$, and $w_+ \propto
\Delta\mathcal{E}$. The physical meaning of this result is as
follows: If the detector moves with constant velocity in the
vacuum of a QFT there is zero probability for it to get excited,
$w_- = 0$. However, if the detector was originally in the excited
state, there is a non--zero probability for it to radiate
spontaneously, $w_+ \neq 0$.

For the case of {\it eternal}, constant, linear acceleration
 -- $x(t) = \left(\frac{1}{a}\sinh\left[a\, t \right], \,
\frac{1}{a}\cosh\left[a\, t \right], \, 0, \, 0\right)$ with $t$
the detector's proper time and $a$ its acceleration -- the
Wightman function is: \bqa G\left[x(t -
\tau/2)^{\phantom{\frac12}},\,\,x(t + \tau/2)\right] \propto
\frac{a^2}{ \sinh^2\left[\frac{a}{2}\, \left(\tau - {\rm i} \,
\epsilon\right)\right]}. \label{lineac} \eqa The integral in
\eq{grule2} is taken using contour integration in the complex
$\tau$ plane. Since $\Delta\mathcal{E} > 0$, the integral $w_-$ in
\eq{grule2} uses a contour which is closed with a large, clockwise
semi-circle in the lower complex half--plane. This contour is
denoted by $C_-$. For $w_+$ the contour is closed with a large,
counterclockwise semi-circle in the upper complex half--plane, and
is denoted by $C_+$. This choice of contours for $w_{\mp}$ is used
everywhere below.

Unlike the constant velocity case, the Wightman function now has
non--trivial poles encircled by the $C_-$ contour, hence, $w_-
\neq 0$. The positions of the poles are easy to find, so the
integral in \eq{grule2} can be calculated exactly with the result:
\bqa w_- \propto
\frac{\Delta\mathcal{E}}{e^{\frac{2\,\pi\,\Delta\mathcal{E}}{a}} -
1}, \quad w_+ \propto \Delta\mathcal{E}\,\left[1 -
\frac{1}{e^{\frac{2\,\pi\,\Delta\mathcal{E}}{a}} - 1}\right].
\label{wwlin} \eqa Therefore a detector moving with constant
acceleration in the vacuum of the background QFT does detect
particles. The detected particles have a Planckian distribution
with temperature $T = \frac{a}{2\, \pi }$ \cite{Unruh}. The
detector gets excited because there is a non--trivial correlation
between field excitations of $\phi$ along its trajectory. The
nontrivial contribution to $w_-$ comes from the non--trivial poles
in the complex $\tau$ plane at $\tau = 2\,\pi\, {\rm i}\, n/a$,
where $n$ is negative integer number. Note that along the
trajectory of a detector fixed at a spatial point in the vicinity
of a Schwarzschild black hole the Wightman function will have the
same analytic features, i.e. the detector will click for the same
physical reason as the accelerating one.

Is it really physically correct to take into account the
contributions of such poles? They are definitely present for
eternal, linear acceleration. However, if one considers a more
realistic linear acceleration with starting/stopping of the
accelerations these initial/final conditions increase the
difficulty of the analysis making it much harder to get a clear
physical picture of what is going on.

Instead of performing a new calculation for a finite time,
linearly, accelerating detector we turn our attention to circular
motion. We will consider homogeneous circular motion, i.e. eternal
circular motion with no starting or stopping. We argue -- via the
specific example where our two--energy level detector is a
electron in an external magnetic field -- that homogeneous
circular motion is a good approximation for real circular motion
with a starting/stopping times. Moreover, in this type of detector
the contribution of the non--trivial poles has been {\it
experimentally verified}.

Now, following \cite{BellLeinaas}, we show that non--trivial poles
appear in the case of a homogeneously orbiting detector
interacting with $\phi$. The trajectory of such a detector with
radius $R$ and angular velocity $\omega_0$, is $x(t) = (\gamma \,
t,\, R\, \cos\left[\gamma \, \omega_0 \, t\right], \, R\,
\sin\left[\gamma \, \omega_0 \, t\right], \,0 )$, $\gamma =
1/\sqrt{1 - R^2 \, \omega_0^2}$ and $t$ is the detector's proper
time.

Inserting this trajectory into \eq{explic}, we obtain: \bqa
G\left[x(t - \tau/2)^{\phantom{\frac12}},\,\,x(t + \tau/2)\right]
\propto \frac{1}{\left[\gamma\, \left(\tau - {\rm i}\,
\epsilon\right)\right]^2 - 4\, R^2 \, \sin^2\left[\frac{\gamma \,
\omega_0}{2}\, \tau\right]}. \label{cirkmot} \eqa This two--point
correlation function has poles in the lower complex $\tau$ plane
enclosed by $C_-$. These poles are similar in nature to those of
the Wightman function for a heat bath \cite{AkhSing3} or for
linear acceleration \eq{lineac}, which lead to a Boltzmann type
exponential contribution to $w_{\mp}$.

For the case of circular motion the velocity is $v = \omega_0\,
R\,\gamma$ and the acceleration is $a = \gamma^2\, \omega_0^2 \,
R$ in the instantaneously, co--moving inertial frame. Unlike the
case of eternal, linear acceleration the integral in \eq{grule2}
for $w_{\mp}$ for the case of orbiting motion can not be done
exactly, since we do not know the exact position of all the poles
in \eq{cirkmot}. However, assuming that the energy splitting is
not too small (i.e. $\Delta\mathcal{E} > a$) we can approximately
find the probability rate \cite{BellLeinaas}: \bqa w_- \propto a
\, e^{- \sqrt{12}\, \frac{\Delta\mathcal{E}}{a}}, \quad w_+
\propto a \, \left(e^{- \sqrt{12}\, \frac{\Delta\mathcal{E}}{a}} +
4\, \sqrt{3} \,\frac{\Delta\mathcal{E}}{a} \right).\label{wwpm}
\eqa The exponential contributions come from the non--trivial
poles in (\ref{cirkmot}) at $\tau \approx \pm {\rm i}\,
\sqrt{12}/a$. The non--exponential contribution to $w_+$ comes
from the trivial pole at $\tau = {\rm i} \, \epsilon$, and is
present even if $a=0$, i.e. corresponds to spontaneous radiation.

Whereas \eq{wwlin} implies a thermal spectrum for linear
acceleration, the results of \eq{wwpm} show that the spectrum
observed by an orbiting detector is not thermal. Intuition from
condensed matter informs us that the Planckian distribution is
strongly related to the form of the two--point correlation
function  in \eq{lineac}. The two--point function for circular
motion, given in \eq{cirkmot}, has a drastically different form
from that in \eq{lineac}.

Thus, we see that the circular Unruh effect has the same physical
origin as the linear case: detectors in homogeneous motion get
excited due to non--trivial correlations between field
fluctuations along their trajectories. Now we are going to show
that the circular Unruh effect has been well known for a long time
but under the name ``Sokolov--Ternov effect". Since the
Sokolov--Ternov effect is experimentally verified this shows that
the non--trivial poles are not  simply a mathematical abstraction,
but have a physical meaning.

Interestingly the same Wightman function just investigated for the
orbiting observer appears in the calculation of the
Sokolov--Ternov effect \cite{SokTer}. This is not a coincidence.
See in particular the derivation of the Sokolov--Ternov effect in
\cite{AkhSing3} \cite{LL4}. We repeat the main steps of this
calculation, but for an arbitrary gyromagnetic number $g$.

The Sokolov--Ternov effect describes the partial depolarization of
electrons in a magnetic field in storage rings due to synchrotron
radiation. It is well known that electrons in circular motion
radiate due to their charge. Apart from this electrons have two
energy levels in an external constant magnetic field bending their
trajectories: with their spins along or against the direction of
the magnetic field. Hence, they can also radiate via flips of
their spins. This spin flip radiation is strongly suppressed in
comparison with the radiation due to the electric charge
\cite{Jackson}.

At first it seems that the spin flip radiation should eventually
polarize the electron beam completely. However, the flips can
happen in both ``directions'' --- either decreasing or increasing
the spin energy. Due to the latter effect the polarization is not
complete. To understand the relation of this effect to the Unruh
effect let us, first, note that electrons can be considered as
quasi-classical detectors (such as the Unruh--DeWitt detector with
two energy levels) when they move ultra--relativistically. In this
case we can neglect both quantization of their motion and
back-reaction to the photon radiation. Apart from this in the
non--inertial, co--moving reference frame the electrons are at
rest. The spin flip transition which decreases the spin energy can
happen due to spontaneous radiation. But what is the reason for
the spin flip transition which increases the spin energy in this
frame where the electrons are at rest? We will show that the
latter transition happens due to existence of the universal
radiation in the non--inertial co--moving reference frame, i.e.
for the same physical reason as in the case of the Unruh effect
appearing for the detector interacting with $\phi$. Posed another
way --- the effect appears due to the non--trivial field
correlations along the orbiting trajectory of the electrons.

The probability rate of synchrotron radiation from a spin flip,
can be obtained from the relativistic equation of motion for a
spin $\vec{s}$ as given by \cite{LL4}: \bqa \frac{d \vec{s}}{d t}
& = & {\rm i} \, \left[\hat{H}_{int}, \,
\vec{s}\right], \nonumber \\
\hat{H}_{int} & = & - \frac{e}{m} \, \vec{s}\, \left[\left(\alpha
+ \frac{1}{\gamma}\right)\, \vec{H} - \frac{\alpha\,
\gamma}{\gamma + 1}\, \vec{v}\,\left(\vec{v} \cdot \vec{H}\right)
 - \left(\alpha + \frac{1}{\gamma + 1}\right)\, \vec{v} \times
\vec{E}\right],\label{hinter} \eqa where $t$ is now the laboratory
time, $\alpha = (g-2)/2$, $\vec{v}$ is the particle's velocity,
$\gamma = 1/\sqrt{1 - v^2}$ and $\vec{E}$ and $\vec{H}$ are the
electric and magnetic fields. Using the interaction Hamiltonian
from \eq{hinter} we can derive the probability rates for photon
emission with spin flips \cite{AkhSing3}: \bqa w_{\mp} \propto
\oint_{C_{\mp}} d\tau\, e^{\mp {\rm i} \, \omega_s \, \tau}\,
\hat{W} \, \left.\frac{1}{ \left(\tau - {\rm i}\,
\epsilon\right)^2 - \left( \vec{r} - \vec{r}'\right)^2}\right|_{r
= r\left(t - \frac{\tau}{2}\right), \,\, r' = r\left(t +
\frac{\tau}{2}\right)}. \label{integralevent} \eqa $\hat{W}$
\cite{AkhSing3} is a differential operator acting on $t$ and $r$.
It appears due to the fact that our ``detectors'' interact with
the electric and magnetic fields rather than directly with the
vector--potential (see \eq{hinter}).

Now in \eq{integralevent} we insert for $r(t)$ a homogeneous
circular trajectory: $(t, R \, \cos \omega_0 t, R \,\sin \omega_0
t, 0)$ with laboratory time, $t$. We can do this, despite the fact
that the real motion of electrons has starting/stopping points,
because the main contribution to the integral in
\eq{integralevent} comes from very small times $\tau$ (to
understand this point one should examine the alternative
stationary phase calculations of the probability rates $w_{\mp}$
in  \cite{SokTer} and \cite{Jackson}). Thus, in \eq{integralevent}
we have the same Wightman function as in \eq{cirkmot}. Note that
$\Delta \mathcal{E}$ is replaced by $\omega_s = [1 + \gamma \,
(g-2)/2]\,\omega_0$ and $\omega_0 = e\, H_b/\mathcal{E}$ is the
cyclotron frequency of an electron with energy, $\mathcal{E}$, in
constant background magnetic field, $H_b$; $\omega_s$ is the
energy difference between electron's spin states in a constant,
background magnetic field. The differential operator $\hat{W}$ is
the source of the difference between the standard Sokolov--Ternov
and circular Unruh effects for detectors interacting with scalar
fields.

Taking the integral in \eq{integralevent}, and considering only
$\alpha > 0$ yields \cite{AkhSing3} \cite{Jackson} : \bqa w_{\mp}
\approx \frac{5\, \sqrt{3} \, e^2 \gamma^5}{16\,m^2 \, R^3} \,
\left\{F_1(\alpha)\, e^{-\sqrt{12} \,\alpha} + F_2(\alpha) \mp
F_2(\alpha)\right\},\label{FF} \eqa where \bqa F_1(\alpha) &=&
\left(1 + \frac{41}{45}\, \alpha - \frac{23}{18}\, \alpha^2 -
\frac{8}{15}\, \alpha^3 + \frac{14}{15} \,\alpha^4\right) -
\frac{8}{5\,\sqrt{3}}\, \left(1 + \frac{11}{12}\, \alpha -
\frac{17}{12}\, \alpha^2 - \frac{13}{24}
\,\alpha^3 + \alpha^4\right), \nonumber \\
F_2(\alpha) &=& \frac{8}{5\,\sqrt{3}}\, \left(1 + \frac{14}{3}\,
\alpha + 8 \, \alpha^2 + \frac{23}{3} \,\alpha^3 +
\frac{10}{3}\,\alpha^4 + \frac{2}{3}\, \alpha^5\right). \eqa Note
the exponential factor in \eq{FF}, which appears for the same
reason as the one in \eq{wwpm}: in both cases the Wightman
functions have the same pole in the lower complex $\tau$ plane. If
$g=2$ (i.e. $\alpha = 0$) we obtain the standard Sokolov--Ternov
expression: \bqa \label{polar-4} w_{\mp} \approx \frac{5\,
\sqrt{3}}{8}\, \frac{e^2\,\gamma^5}{m^2\, R^3}\, \left(1 \mp
\frac{8\, \sqrt{3}}{15}\right). \eqa In this case the exponent is
equal to 1. This is the reason why the exponential factor, the
hallmark of the Unruh effect, is usually overlooked in the
standard Sokolov--Ternov considerations. Note that the exponential
factor is always present in the form $e^{1/\gamma}$ even if $g$ is
exactly $2$, but we are taking $\gamma \gg 1$. In any case, if we
consider $g \neq 2$, then the exponent is  explicitly present.  In
the case of the Sokolov--Ternov effect we have
$\Delta\mathcal{E}/a \approx (g-2)/2$ if $\gamma\gg 1$. Thus, the
laboratory observer interprets the effect as the Sokolov--Ternov
effect, while the non-inertial co-moving observer interprets the
effect as the circular Unruh effect. Physically these two effects
are the same. The connection between the Unruh and Sokolov-Ternov
effects has been previously discussed in \cite{BellLeinaas}, \cite{Mane}
and \cite{Unruh2}.  \\

{\bf Acknowledgment} AET would like to thank V.Zakharov,
M.Polikarpov, S.Mane, M.Danilov, A.Mironov, A.Morozov and
N.Narozhny for valuable discussions. This work supported by the
CSU Fresno International Activities Grant. AET would like to thank
INTAS 03-51-5460 grant and Agency of Atomic Energy of Russian
Federation for the financial support.

\end{document}